\documentclass{article}

\PassOptionsToPackage{square,comma}{natbib}

\usepackage[final]{neurips_2024}

\usepackage[utf8]{inputenc} 
\usepackage[T1]{fontenc}    
\usepackage{hyperref}       
\usepackage{url}            
\usepackage{booktabs}       
\usepackage{amsfonts}       
\usepackage{nicefrac}       
\usepackage{microtype}      
\usepackage{xcolor}         
\usepackage{amsmath}
\usepackage{booktabs}
\usepackage{graphicx}
\usepackage{multirow}
\usepackage{makecell}

\title{Quantize More, Lose Less: Autoregressive Generation from Residually Quantized Speech Representations}

\author{%
  AMAP Speech
}

\begin{document}

\maketitle

\begin{abstract}
  Text-to-speech (TTS) synthesis has seen renewed progress under the discrete modeling paradigm. Existing autoregressive approaches often rely on single-codebook representations, which suffer from significant information loss. Even with post-hoc refinement techniques such as flow matching, these methods fail to recover fine-grained details (e.g., prosodic nuances, speaker-specific timbres), especially in challenging scenarios like singing voice or music synthesis.
  
We propose QTTS, a novel TTS framework built upon our new audio codec, QDAC. The core innovation of QDAC lies in its end-to-end training of an ASR-based auto-regressive network with a GAN, which achieves superior semantic feature disentanglement for scalable, near-lossless compression. QTTS models these discrete codes using two innovative strategies: the Hierarchical Parallel architecture, which uses a dual-AR structure to model inter-codebook dependencies for higher-quality synthesis, and the Delay Multihead approach, which employs parallelized prediction with a fixed delay to accelerate inference speed.
Our experiments demonstrate that the proposed framework achieves higher synthesis quality and better preserves expressive content compared to baseline. This suggests that scaling up compression via multi-codebook modeling is a promising direction for high-fidelity, general-purpose speech and audio generation.
\end{abstract}

\section{Introduction}
Traditional speech synthesis approaches mostly rely on non-autoregressive architectures(\citet{kim2021conditional}, \citet{ren2022fastspeech2fasthighquality}, \citet{li2023styletts2}). However, recent research has demonstrated that large autoregressive models are more effective at learning accurate alignments between text and audio(\citet{borsos2023audiolm}, \citet{kharitonov2023speakreadprompthighfidelity}, \citet{yang2023uniaudio}). With this advancement, large language model (LLM)-based text-to-speech (TTS) systems have become increasingly mainstream. These systems convert text into sequences of discrete audio tokens, which are subsequently transformed into waveforms using neural vocoders.
In practice, many such systems use a single discrete codebook to quantize semantic features(\citet{betker2023better}, \citet{anastassiou2024seed}, \citet{lajszczak2024base}, \citet{casanova2024xtts}, \citet{zhang2025minimax}, \citet{guo2024fireredtts}). For example, CosyVoice employs one VQ encoder with only 4096 codes(\citet{du2024cosyvoice}). While simple, this bottleneck loses fine acoustic detail. In fact, the CosyVoice 2 authors explicitly note that their single-codebook model “does not perform well when tasked with singing(\citet{du2024cosyvoice2}). In other words, single-codebook models struggle to capture the rich high-frequency content and prosodic nuance required for singing or music synthesis.

To avoid the limitations of single-codebook representations, some works have adopted diffusion-based methods. However, these approaches give rise to a new challenge: the modeling of duration. For instance, F5-TTS pads the text sequence with blank tokens to match the waveform length(\citet{chen2024f5}), and MaskGCT predicts the duration of the entire sentence(\citet{wang2024maskgct}). These methods bypass detailed timing inference but at the cost of fine-grained control. Because they model duration only coarsely, they tend to miss the subtle timing and intonation patterns in natural speech, leading to less precise prosody and expressiveness. Moreover, these methods still require a vocoder to convert mel-spectrograms into waveforms, which introduces additional information loss in the reconstruction process(\citet{kong2020hifigangenerativeadversarialnetworks}, \citet{kumar2019melgan}, \citet{yang2021multi}).

An alternative for audio tokenization is to use multi-codebook residual vector quantization (RVQ) (\citet{zeghidour2021soundstream}, \citet{defossez2022high}). In RVQ, an audio frame is represented by a sum of vectors from several quantizers, allowing for high-fidelity reconstruction over a range of bitrates by capturing details that single-codebook models often miss (\citet{kumar2023high}, \citet{defossez2022high}, \citet{ye2025llasascalingtraintimeinferencetime}, \citet{wang2025spark}, \citet{parker2024scalingtransformerslowbitratehighquality}).
However, a key challenge in RVQ is the entanglement of different feature types (e.g., semantic, acoustic, prosodic) across the codebooks. Many existing approaches can be considered naive, as they either impose no constraints on the intermediate features or apply a distillation loss only to the first quantization layer. We argue that these methods lack an explicit mechanism for feature decoupling, which limits the model's ability to efficiently represent audio.
To address this limitation, we introduce QDAC (Quantization-Decoupled Audio Codec), a novel framework that achieves explicit feature disentanglement. Building upon the foundation of DAC, QDAC uniquely incorporates a pre-trained autoregressive (AR) ASR model to guide the quantization process. By leveraging the strong sequential modeling capabilities of the AR-ASR model, QDAC effectively decouples semantic information from other acoustic attributes. This stands in contrast to prior single-codebook methods that may also use ASR for decoupling, as our approach deeply integrates the AR model's supervision into a multi-stage RVQ setup for a more robust separation. To manage the trade-off between audio fidelity and inference speed within this framework, we designed two distinct generation architectures.


For the higher audio quality, our Hierarchical Parallel architecture utilizes a dual-autoregressive structure. Specifically, the model predicts the tokens for each codebook layer sequentially, where the generation of the current layer is conditioned on the complete, finalized tokens of all preceding layers. This strict, hierarchical conditioning ensures that crucial inter-codebook dependencies are fully captured, maximizing reconstruction fidelity.
To accelerate synthesis, our Delay Multihead architecture employs a parallelized prediction mechanism with a fixed-step delay. Instead of waiting for an entire codebook layer to be generated, the model begins predicting tokens for the next layer after only a small, fixed number of steps (k steps) of the current layer. This approach breaks strict sequential dependence, allowing for significant computational parallelism while still retaining essential local context between codebooks, a flaw in some fully parallel designs (\citet{copet2023simple}).
Together, these innovations allow QTTS to synthesize extremely high-fidelity audio efficiently. Empirically, we find that QTTS vastly outperforms prior single-codebook and coarse-duration models in objective and subjective quality: it achieves higher audio naturalness and zero-shot speaker similarity, and significantly better MOS ratings. These results demonstrate the advantage of explicitly modeling compression in TTS by preserving fine acoustic details, QTTS delivers richer, more accurate speech.

The architecture of QTTS is built upon the following key contributions:
\begin{enumerate}
\item \textbf{A semantically-aware audio codec, QDAC.} We propose QDAC, which moves beyond naive compression. To achieve this, QDAC leverages a AR-based ASR model to explicitly guide the first codebook of our vector quantizer to capture semantic information (e.g., phonemes). This ensures that subsequent codebooks can focus on modeling finer, residual acoustic details, leading to a more efficient and meaningful representation.

\item \textbf{A choice of specialized decoding strategies.} To effectively model these discrete tokens, we introduce two complementary autoregressive strategies, providing users a direct trade-off between synthesis quality and inference speed:
\begin{itemize}
\item \textbf{Hierarchy Parallel for High-Fidelity Synthesis:} This strategy employs a dual-autoregressive structure to model the full hierarchical dependency between codebooks. By ensuring each codebook is generated with complete contextual awareness of all preceding ones, it maximizes audio quality and naturalness.
\item \textbf{Multihead Delay for Fast Inference:} This strategy parallelizes the generation process. It uses a multi-head mechanism to predict subsequent codebooks with a fixed-step delay, significantly reducing sequential dependencies and accelerating inference speed for real-time applications.
\end{itemize}
\end{enumerate}



\section{Method}
\subsection{Audio Tokenizer}
\begin{figure}[htbp]
  \centering
  \includegraphics[width=1.0\textwidth]{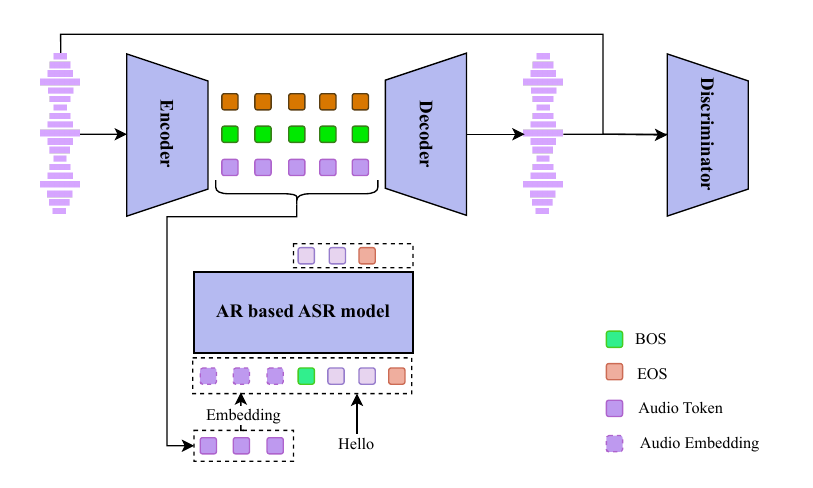}
  \caption{Architecture of the audio tokenizer model. It adopts a GAN-based framework consisting of an encoder-decoder generator and a discriminator, trained with semantic guidance from a AR based ASR model. During training, only the first codebook is explicitly supervised by semantic loss, while the remaining codebooks are left to learn freely, allowing the model to flexibly capture additional residual details.}
  \label{fig:graph1}
\end{figure}
\subsubsection{Residual Vector Quantization}
For each input frame t, an encoder produces a continuous latent vector $\mathbf{f}_t \in \mathbb{R}^D$. 
We quantize this vector in C sequential stages, where each stage refines the residual error left by the previous one:
$$
\begin{aligned}
r_t^{(0)} &= f_t, \\
c_t^{(i)} &= \arg\min_{k} \big\| r_t^{(i-1)} - e_k^{(i)} \big\|, \\
r_t^{(i)} &= r_t^{(i-1)} - e_{c_t^{(i)}}^{(i)}, \\
\hat{f}_t &= \sum_{i=1}^{C} e_{c_t^{(i)}}^{(i)},
\end{aligned}
$$
Here, $e_k^{(i)} \in \mathbb{R}^D$ refers to the k-th prototype vector in the $i$-th codebook, and $\hat{f}_t$ is the reconstructed latent after all quantization stages. Each chosen index $c_t^{(i)}$ is then mapped to a 1024-dimensional embedding via $W^{(i)} \in \mathbb{R}^{1024 \times 2048}$. This cascaded design enables progressively finer approximation of $\mathbf{f}_t$, an approach similar to that adopted in state-of-the-art audio codecs like Encodec and SoundStream.
\subsubsection{Disentangling Information}

Prior approaches, such as MiMi (\citet{defossez2024moshi}), often rely on distillation from general-purpose Self-Supervised Learning (SSL) models like WavLM(\citet{Chen_2022}). While powerful, these SSL models are not explicitly optimized for disentanglement. Consequently, their learned representations tend to entangle crucial semantic content with other acoustic attributes like speaker identity, prosody, and recording artifacts, limiting their effectiveness for fine-grained generation and editing tasks.

To address this limitation, our codec, QDAC, is specifically designed to enforce a clear separation of information across its quantized codebooks. As show in Figure \ref{fig:graph1}, the core objective is to isolate linguistic information into a designated primary codebook, compelling the subsequent codebooks to capture the remaining residual acoustic information.
We achieve this via a multi-objective training strategy that operates end-to-end. The process involves two simultaneous tasks with distinct objectives and information pathways:
Waveform Reconstruction: A decoder, trained adversarially within a GAN framework, is tasked with reconstructing the original waveform. This decoder receives tokens from all codebooks $C_{1...N}$, and its associated reconstruction and adversarial losses ensure that the combined representation is sufficient for high-fidelity, perceptually rich audio synthesis.Semantic Disentanglement: Concurrently, an autoregressive ASR module is trained to predict the text transcription. Crucially, this ASR module is conditioned exclusively on the tokens from the first codebook $C_1$. The cross-entropy loss from this ASR prediction is therefore backpropagated only through this primary codebook. This acts as a powerful supervisory signal, forcing $C_1$
 to encode the semantic content necessary for transcription, while leaving the other codebooks $C_
{2...N}$ free to be optimized by the reconstruction loss for capturing non-semantic information like speaker timbre and prosody.
\begin{figure}[htbp]
  \centering
  \includegraphics[width=1.0\textwidth]{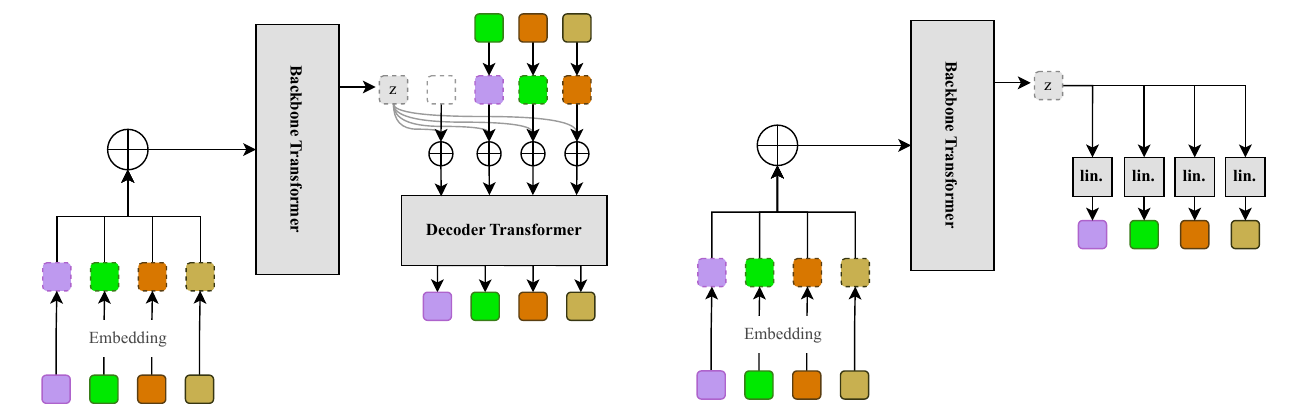}
  \caption{Comparison of two multi-codebook prediction strategies for TTS: (Left) The Parallel Hierarchy structure decodes tokens autoregressively across both time and codebook dimensions, allowing full contextual conditioning at the cost of slower sequential inference. (Right) The Multihead-Delay architecture predicts each codebook stream with a dedicated head and fixed delay offset, enabling parallel inference but with limited cross-codebook context.  Each approach represents a trade-off between modeling capacity and runtime efficiency.}
  \label{fig:graph2}
\end{figure}
\subsection{Multi-Head Delay-Pattern Prediction}
In this approach, the model assigns a separate prediction head to each of the K codebooks, enabling parallel decoding of the multi-codebook output. All heads share a common encoder representation, but each head is trained to align its output with a fixed temporal offset or delay relative to the others. This delay pattern reflects the intuition that RVQ codebooks are hierarchical: the first (coarsest) codebook carries the main acoustic content, and each subsequent codebook encodes residual fine detail conditioned on the earlier ones. Concretely, the decoding for codebook k is shifted by k time steps so that codebook k at time t can depend on codebook (k-1) at the same frame.  
In practice, the multi-head structure enables parallel decoding where all K heads emit their tokens simultaneously at each time frame. This is achieved by combining one head per codebook with a structured delay scheme via specific input offsets. This design efficiently captures codebook-wise dependencies, as each head's output is causally conditioned on necessary lower-level predictions. In addition, the delay-parttern scheme enables each codebook's stream to learn its own unique alignment delay and temporal dynamics. Crucially, the overall prediction remains autoregressive to the acoustic target, strictly preserving the causal nature vital for high-quality speech synthesis.

While the multi-head delay strategy significantly accelerates inference, it involves a deliberate trade-off in contextual modeling. Because each predictive head operates with a fixed delay, the model generates tokens at each step without access to the complete context from all preceding codebooks. This restricted receptive field across the codebook dimension can limit the model's ability to capture the most complex, long-range dependencies. Consequently, this approach prioritizes a major gain in synthesis speed at the cost of the exhaustive contextual modeling found in fully autoregressive structures, which may affect the highest level of reconstruction fidelity.

To mitigate this limitation, one straightforward approach is to extend the delay range for each codebook, granting higher-level heads access to a broader historical context of preceding tokens. This direct increase in receptive field aims to indirectly incorporate more semantic and structural cues from earlier codebooks. However, increasing the delay length introduces its own challenges. long-range dependencies become harder to model effectively due to the causal links within hierarchical RVQ codebooks. Furthermore, the model may struggle to align delayed inputs with the current generation context. Thus, while multihead-delay offers compelling efficiency gains, it must balance delay depth and modeling capacity to avoid sacrificing audio quality.

\subsection{Hierarchy Parallel Prediction}
The dual auto-regressive (AR) RQ Transformer models the residual vector quantization (RVQ) output as a two-level autoregressive process, operating first along the temporal axis and subsequently across codebooks. As show in Figure \ref{fig:graph2}, the core intuition behind this design is to preserve both the causal nature of speech generation and the hierarchical refinement characteristic of RVQ. Specifically, the model predicts quantized acoustic tokens frame by frame, proceeding autoregressively in time. At each time step t, the model generates a sequence of K discrete tokens corresponding to the RVQ codebooks, in a coarse-to-fine order. This inner sequence is itself autoregressive across codebook indices: the token from codebook k is conditioned not only on the tokens from all preceding time steps, but also on the codebook predictions \{0, 1, ..., k-1\} at the current frame t.
This dual-AR RQ Transformer architecture leverages a hierarchical decomposition to first construct a rough acoustic target using the top-level codebook. It then incrementally refines this representation with finer-grained residuals from deeper codebooks. The model enforces temporal continuity through an outer AR loop over time, ensuring causality and coherent speech progression. Meanwhile, the inner AR path over codebooks captures inter-level dependencies, effectively modeling how fine-scale details emerge as corrections to previously predicted residuals. Architecturally, this approach integrates both dimensions within a unified decoding loop, where the transformer processes one frame at a time, generating codebooks sequentially and leveraging past temporal context. By explicitly encoding both intra-frame quantization hierarchy and inter-frame dynamics, the dual-AR RQ Transformer provides a principled and scalable solution for multi-codebook token generation in large-scale TTS models.

The parallel hierarchy approach addresses the limitations of contextual visibility inherent in delay-based multihead architectures. It models token generation process as a two-level structure: first over time, then over codebook depth. This ensures that each prediction step has access to a full and coherent context, both temporally and hierarchically. Unlike multihead-delay methods, it does not rely on staggered input streams or fixed delay offsets. Instead, it preserves a clean autoregressive dependency structure where each frame is decoded in sequence, and each codebook within a frame is conditioned on all previously predicted levels. This allows the model to capture subtle inter-codebook dependencies and maintain high fidelity across both coarse and fine acoustic features.

The primary cost of this method lies in its inference speed. Since the decoding process is sequential in two dimensions (time and codebook), it cannot achieve the same degree of parallelism as delay-based methods. However, the input sequence fed to the transformer backbone is significantly shorter than in flattened representations, which contributes to better stability and convergence during inference. Moreover, modern inference frameworks often support decoding with fixed-length segments, making it possible to apply parallelism across batched codebook layers. This structured execution can lead to practical runtime acceleration and a favorable real-time factor (RTF), even for autoregressive decoders. Consequently, while the parallel hierarchy model sacrifices some raw speed, it benefits from improved modeling accuracy and stability, making it a competitive choice in high-quality TTS systems.

\section{Experiments}

\subsection{Inference Acceleration and Performance Benchmarking}
In this section, we present inference acceleration strategies for the proposed Hierarchy Autoregressive (AR) structure and benchmark our approach in terms of real-time performance and throughput. Specifically, we evaluate the highest achievable real-time factor and throughput of our model.
To enable the application of popular large language model (LLM) inference acceleration frameworks (e.g., vLLM, SGlang) to our model architecture, we modified the vLLM library to accommodate the unique requirements of our AR inference design. AR inference is inherently dynamic, as the termination point of the AR process—i.e., the number of tokens generated—cannot be predetermined. For the Backbone AR, we leveraged vLLM's PagedAttention to manage the key-value (KV) cache efficiently. To avoid the complexity of simultaneously managing two dynamic AR inference processes, we reformulated the inference process for the Decoder AR into a static inference setup.
For the Decoder AR, the fixed inference step size can be explicitly defined based on the value of K(the number of codebook tokens). Assuming a Transformer architecture comprising Nblocks, a single decoder inference step involves processing through all Nblocks. Consequently, Ksteps will traverse $ K \times N $ blocks in total, representing a stacked computation graph. We define the Nblocks processed in a single step as a Stacked Block, while the Kstacked steps collectively form a Stacked Decoder Transformer.
To manage the KV cache for the Decoder Transformer, we pre-allocated GPU memory for a KV Cache Block with a fixed sequence length of K. By doing so, we converted the unpredictable sequence length of dynamic inference into a static-dimension process. This allows the Stacked Decoder Transformer to be compiled as a static full computation graph, benefiting from CUDA Graphs to minimize the frequency of CUDA kernel launches and significantly improve overall computational efficiency.
Due to the unique hierarchical architecture of our model, only the Backbone AR is required to perform forward computation during the Prefill phase, while the Decoder AR remains inactive until the Decode phase. This design dramatically reduces the computation overhead in the Prefill phase. Specifically, the Time-To-First-Token (TTFT) corresponds exclusively to the computation time of the Backbone AR during the Prefill phase. Even under long prompts, our model maintains an exceptionally low TTFT and achieves high throughput.

\begin{table}[!ht]
    \centering
    \label{tab:ttft}
    \caption{Time-To-First-Token (TTFT) and Tokens-Per-Output-Time (TPOT)}
    \begin{tabular}{lccc}
        \toprule
        &Backbone Input Length &TTFT (ms) &TPOT (ms)\\
        \midrule
        \multirow{3}{*}{QTTS Hierarchy 200M}
         &512 &26 &5   \\
         &64 &7 &6   \\
         &16 &7 &6  \\
         \midrule
         \multirow{3}{*}{QTTS Multihead 120M}
         &512 &24 &5 \\
         &64 &4 &5 \\
         &16 &4 &5 \\
        \bottomrule
    \end{tabular}
\end{table}

\begin{table}[!ht]
    \centering
    \label{tab:latency}
    \caption{Latency results for the First Chunk}
    \resizebox{\textwidth}{!}{  
    \begin{tabular}{lcccccc}
        \toprule
        &Backbone Input Length &Average RT (ms) &Min RT (ms) &P95 RT (ms) &P99 RT (ms) &Max RT (ms) \\
        \midrule
        \multirow{6}{*}{QTTS Hierarchy 200M}
         &512 &139 &32 &165 &172 &228  \\
         &256 &133 &24 &154 &159 &185  \\
         &128 &130 &24 &149 &155 &179  \\
         &64 &132 &24 &150 &155 &179  \\
         &32 &126 &23 &152 &158 &172  \\
         &16 &108 &22 &149 &156 &169  \\
         \midrule
         \multirow{6}{*}{QTTS Multihead 120M}
         &512 &79 &70 &256 &69 &64 \\
         &128 &68 &64 &64 &68 &63 \\
         &32 &68 &63 &16 &68 &63 \\
         &90 &97 &125 &76 &82 &113 \\
         &75 &82 &125 &74 &81 &125 \\
         &74 &81 &126 &73 &80 &125 \\
        \bottomrule
    \end{tabular}
    }
\end{table}

\begin{table}[!ht]
    \centering
    \label{tab:hierarchy}
    \caption{Token processing speed of QTTS-Hierarchy-200M.}
    \begin{tabular}{llccc}
        \toprule
         & \makecell{Input Length\\(tokens)} & \makecell{Output Length\\(tokens)} & \makecell{Backbone Input\\(tokens/s)} & \makecell{Codebook Output\\(tokens/s)} \\
        \midrule
        \multirow{3}{*}{Prefill}
            & 2048 & 1   & 98{,}303.66   & -- \\
            & 1024 & 1   & 100{,}793.02  & -- \\
            & 128  & 1   & 98{,}612.96   & -- \\
        \midrule
        \multirow{3}{*}{Decode}
            & 1    & 2048 & --           & 74{,}285.20 \\
            & 1    & 1024 & --           & 90{,}401.20 \\
            & 1    & 128  & --           & 105{,}643.76 \\
        \bottomrule
    \end{tabular}
\end{table}

\begin{table}[!ht]
    \centering
    \label{tab:multihead}
    \caption{Token processing speed of QTTS-Multihead-120M.}
    \begin{tabular}{llccc}
        \toprule
         & \makecell{Input Length\\(tokens)} & \makecell{Output Length\\(tokens)} & \makecell{Backbone Input\\(tokens/s)} & \makecell{Codebook Output\\(tokens/s)} \\
        \midrule
        \multirow{3}{*}{Prefill}
            & 2048 & 1   & 91259.8   & -- \\
            & 1024 & 1   & 92951.08  & -- \\
            & 128  & 1   & 85827.98   & -- \\
        \midrule
        \multirow{3}{*}{Decode}
            & 1    & 2048 & --           & 117338.32 \\
            & 1    & 1024 & --           & 148006.24 \\
            & 1    & 128  & --           & 196749.2 \\
        \bottomrule
    \end{tabular}
\end{table}

\subsubsection{Experimental Setup and Results}
We conducted experiments on a single Nvidia H20 GPU with 96GB of memory under fp16 precision. The experiments utilized two versions of the QTTS model: a Hierarchy-structured model with a parameter size of 200M, and a Multihead-structured model with a parameter size of 120M. For all experiments, the codebook VQ was configured with a frequency of 25Hzand a total of 8codebooks.

The Backbone AR module does not directly process input or output tokens. Instead, it takes embedding vectors as input and generates hidden states as output. To simplify performance representation, we treat the input embeddings as the backbone input tokens and the generated hidden states as the backbone output tokens. Correspondingly, the tokens produced by the Decoder AR module are referred to as codebook tokens.

Using backbone tokens as the unit of measurement, we conducted experiments over 1,000 iterations. In the experiments, the first chunk was set to contain 200 codebook tokens (corresponding to 25 backbone tokens, representing 1 second of synthesized audio), with generation performed on a single thread. The latency results for the First Chunk are summarized in Table \ref{tab:latency}. Additionally, we measured the model’s Time-To-First-Token (TTFT) and Tokens-Per-Output-Time (TPOT), with results presented in Table \ref{tab:ttft}.

Throughput is a significant strength of the QTTS model. To assess throughput under different configurations, we performed stress tests across varying input token lengths, output token lengths, and batch sizes. Performance was evaluated independently for both the Prefill and Decode phases. Each experiment was conducted for 10 minutes, and the throughput was calculated as the average value over the duration of the test.

\subsection{Audio Quantization and Reconstruction Analysis}

To assess the compression capacity of QTTS, we study how increasing the number of codebooks improves audio reconstruction quality.

\textbf{Setup.} We use a DAC-style audio tokenizer with 1, 4, 8, and 16 codebooks (each with 2048 entries and 1024-dimensional embeddings), operating at a 25 Hz frame rate. We evaluate zero-shot capabilities on  SeedTTS-Easy and PGC-hard test sets. The SeedTTS-Easy set consists of relatively standard utterances. The PGC-hard set, containing Professionally-Generated Content, comprising 60 challenging vocal IPs, is specifically designed to test model generalization on difficult, out-of-domain voices. It features samples from professional voice actors with highly expressive prosody and speakers with unique timbres, both of which are known to be challenging for zero-shot TTS.

\textbf{Baselines.} To assess the reconstruction quality of our codec, we compare it with the original Descript Audio Codec (DAC) and Speech Tokenizer. For the zero-shot TTS task, we benchmark our model against CosyVoice v1 and CosyVoice v2, two prominent and high-performing open-source baselines.









\textbf{Results.} The results presented in Table \ref{tab:codec} lead to the following key observations: 1. Increasing the number of codebooks yields significantly better reconstruction quality. Our 16-codebook setup achieves near-lossless audio reconstruction, outperforming all previous VQ-based approaches. 2. For a fixed number of codebooks, employing a higher token sampling rate yields marked improvements across all reconstruction metrics. This is evidenced by significantly higher PESQ, STOI, and SI-SDR scores, coupled with lower STFT and Mel distances. 3. The 32-codebook original DAC baseline, operating at the highest bitrate, predictably achieves the best reconstruction results.
\begin{table}[!ht]
    \centering
    \caption{Reconstruction quality of audio codec.}
    \label{tab:codec}
    \resizebox{\textwidth}{!}{  
    \begin{tabular}{@{}lcccccccccc@{}}
        \toprule
        Tokenizer & CB & FR & BR & PESQ $\uparrow$ & STOI $\uparrow$ & SI-SDR $\uparrow$ & STFT $\downarrow$ & Mel $\downarrow$ & WER $\downarrow$ & Spk Sim $\uparrow$ \\
        \midrule
        Ground Truth       & -  & -     & -   & 4.64 & 1.00 & 10.56 & 0.00 & 0.00 & 6.01 & 0.76 \\
        \midrule
        QDAC           & 8  & 25Hz & ~   & 2.98 & 0.94 & 7.70  & 0.17 & 3.62 & 6.42 & 0.90 \\
                           & 8  & 50Hz & 2.2kbps   & 3.07 & 0.94 & 7.68  & 0.11 & 3.36 & 6.38 & 0.91 \\
                           & 16 & 25Hz & 4.4kbps   & 3.80 & 0.93 & 7.72  & 0.11 & 2.99 & 6.24 & 0.93 \\
                           & 16 & 50Hz & 4.4bos   & 3.83 & 0.97 & \textbf{11.77} & 0.08 & 2.55 & \textbf{5.22} & 0.93 \\
        \midrule
        DAC          & 32 & 50Hz & 8.8kbps   & \textbf{4.35} & 0.99 & 10.02 & \textbf{0.03} & \textbf{1.99} & 6.19 & \textbf{0.97} \\
        SpeechTokenizer    & 8  & 66Hz & 16kbps   & 1.26 & 0.77 & 6.66  & 0.58 & 7.02 & 10.86 & 0.68 \\
        \bottomrule
    \end{tabular}
    }
\end{table}

Table \ref{tab:performance_comparison} compares the zero-shot synthesis performance of our model against the CosyVoice v1 and v2 baselines. Our 16-codebook model substantially surpasses the single-codebook baselines (CosyVoice v1/v2) on all metrics. This underscores a fundamental limitation of single-codebook architectures for zero-shot TTS. Furthermore, our 16-codebook model achieves superior performance in WER, Speaker Similarity, and MOS. This finding is consistent with the codec reconstruction evaluation in Table \ref{tab:codec}, thus validating the critical role of high-fidelity codec performance in driving downstream synthesis quality.

\begin{table}[!ht]
    \centering
    \caption{Speech Synthesis Performance}
    \label{tab:performance_comparison}
    \resizebox{\textwidth}{!}{
    \begin{tabular}{@{}llcccccc@{}}
        \toprule
        \textbf{Model} & \textbf{Codebooks} & \textbf{Fr} & \textbf{Model Size} & \textbf{WER↓} & \textbf{PGChard} & \textbf{Spk Sim↑} & \textbf{MOS↑} \\
        \midrule
        Ground Truth & - & - & - & 1.25 & 6.01 & 0.76 & 2.70 \\
        \midrule
        QTTS & 8 & 25 Hz & 0.2B & 1.66 & 6.89 & 0.75 & 3.03 \\
        CosyVoice & 1 & 50 Hz & 0.3B & 3.05 & 7.86 & 0.82 & 3.01 \\
        CosyVoice2 & 1 & 25 Hz & 0.5B & 1.32 & 6.11 & 0.81 & 3.02 \\
        \bottomrule
    \end{tabular}
    }
\end{table}

\section{Conclusion}
In this work, we presented QTTS, a compression-aware TTS framework that leverages multi-codebook residual vector quantization (RVQ) for high-fidelity speech generation. By scaling up the number of codebooks and applying feature disentanglement, our discrete audio tokenizer achieves near-lossless reconstruction while preserving fine-grained acoustic and prosodic details. To effectively model these rich representations, we introduced two autoregressive decoding strategies: Multihead Delay, which accelerates inference through parallel codebook prediction, and Hierarchy Parallel, which improves synthesis quality by preserving inter-codebook dependencies.
Empirical results demonstrate that QTTS significantly outperforms previous single-codebook and coarse-duration models across a range of speech tasks, including zero-shot speaker transfer and expressive synthesis. These findings underscore the importance of explicitly modeling compression in TTS and highlight multi-codebook architectures as a promising direction for scalable, high-quality speech and audio generation.

Future work may explore extending QTTS to multi-modal scenarios (e.g., singing voice, music, or cross-lingual TTS), as well as investigating non-autoregressive alternatives for further improving inference speed without sacrificing quality.

\clearpage
\bibliographystyle{plainnat}
\bibliography{main}

\clearpage
\appendix
\section{Contributors}
Core Contributors:
Yichen Han, Xiaoyang Hao, Keming Chen, Weibo Xiong, Jun He, Ruonan Zhang, Junjie Cao, Yue Liu$^{*}$\footnote{* Corresponding Author.}

Contributors: Bowen Li, Dongrui Zhang, Hui Xia, Huilei Fu, Kai Jia, Kaixuan Guo, Mingli Jin, Qingyun Meng, Ruidong Ma, Ruiqian Fang, Shaotong Guo, Xuhui Li, Yang Xiang, Ying Zhang, Yulong Liu, Yunfeng Li, Yuyi Zhang, Yuze Zhou, Zhen Wang, Zhaowen Chen




\end{document}